\documentclass[a4paper,11pt]{article}
\usepackage{pos}
\usepackage{graphicx}
\usepackage{slashed}
\usepackage{tikz}
\usepackage{tikz-feynman}
\tikzfeynmanset{compat=1.0.0}

\newcommand{\csw}{{c_\text{SW}}}
\newcommand{\cswzero}{{{\csw}^{(0)}}}
\newcommand{\cswone}{{{\csw}^{(1)}}}
\newcommand{\intdk}{\int\limits_{-\pi}^\pi\frac{\mathrm{d}^4k}{(2\pi)^4}}

\title{Stout-smearing, gradient flow  and $c_{\text{SW}}$ at one loop order}

\author*[a]{Maximilian Ammer}
\author[a,b]{Stephan Dürr}

\affiliation[a]{Physics Department, University of Wuppertal, D-42119 Wuppertal, Germany}

\affiliation[b]{IAS/JSC, Forschungszentrum Jülich, D-52425 Jülich, Germany}

\emailAdd{ammer\,(AT)\,uni-wuppertal$\mbox{.}$.de}
\emailAdd{duerr\,(AT)\,uni-wuppertal$\mbox{.}$de}

\abstract{The one-loop determination of the coefficient $\csw$ of the Wilson quark
action has been useful to push the leading cut-off effects for on-shell
quantities to $\mathcal{O}(\alpha^2 a)$ and, in conjunction with non-perturbative
determinations of $\csw$, to $\mathcal{O}(a^2)$, as long as no link-smearing is
employed.
These days it is common practice to include some overall link-smearing
into the definition of the fermion action. Unfortunately, in this
situation only the tree-level value $c_\text{SW}^{(0)}=1$ is known, and
cut-off effects start at $\mathcal{O}(\alpha a)$. We present some general techniques
for calculating one loop quantities in lattice perturbation theory
which continue to be useful for smeared-link fermion actions.
Specifically, we discuss the application to the 1-loop improvement
coefficient $c_\text{SW}^{(1)}$ for overall stout-smeared Wilson fermions.}

\FullConference{%
 The 38th International Symposium on Lattice Field Theory, LATTICE2021
  26th-30th July, 2021
  Zoom/Gather@Massachusetts Institute of Technology
}

\begin{document}
\maketitle

\section{Introduction}
In the original work  Sheikholeslami and Wohlert \cite{Sheikholeslami:1985ij} noted that the improvement coefficient $\csw$ has to be equal to one at tree level. Wohlert later calculated its value at one loop order using twisted boundary conditions as a regulator \cite{Wohlert:1987rf}. More one-loop calculations of $\csw$ have been performed by Aoki and Kuramashi \cite{Aoki:2003sj} focusing on different improved gauge actions and by Horsley et al. \cite{Horsley:2008ap} for SLiNC fermions (i.e. with stout-smearing in the Wilson part of the fermion action but not in the clover term). Our aim is to calculate $\csw$ to one loop order for the Wilson-clover action including overall stout smearing. Furthermore because of the close relation of stout-smearing to the gradient flow of the Wilson action we are able to directly extend our results to the case of flowed fields.

\section{The perturbative determination of $\csw$\label{sect:2}}
The Sheikholeslami-Wohlert-coefficient  $\csw$ of the $\mathcal{O}(a)$-improved action:
\begin{align}
\mathcal{S}_I=\mathcal{S}_\text{Wilson}+c_\text{SW}\cdot \sum\limits_x\sum\limits_{\mu<\nu}\bar{\psi}(x)\frac 12 \sigma_{\mu\nu}F_{\mu\nu}(x)\psi(x)
\end{align}
has a perturbative expansion  $c_\text{SW}=c_\text{SW}^{(0)}+g_0^2c_\text{SW}^{(1)}+\mathcal{O}(g_0^4)$ in powers of the bare coupling $g_0^2=2N_c/\beta$. It can be calculated via the quark-quark-gluon-vertex function 

\begin{align}
\includegraphics[scale=0.2]{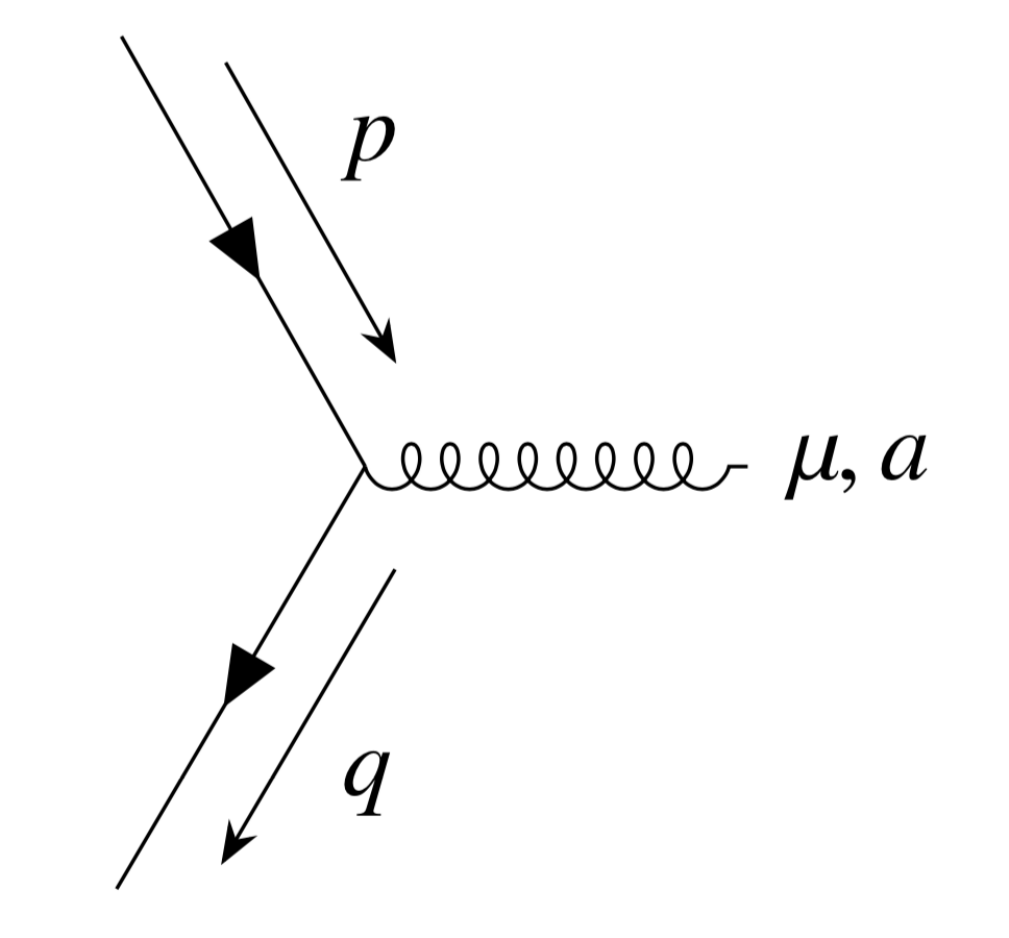}
\qquad 
\Lambda^a_\mu(p,q)=\sum\limits_{L=0}^\infty g_0^{2L+1} \Lambda^{a(L)}_\mu(p,q).
\end{align}
The number of loops $L$ corresponds to the numbering of the expansion coefficients $c_\text{SW}^{(L)}$.
At tree level it is given by the lattice version of the qqg-vertex:
\begin{align}
 &\Lambda^{a(0)}_\mu(p,q)=(V_1^a)_\mu(p,q)\\
& =-g_0 T^a\left(i\gamma_\mu + a\left(\frac{1}{2}(p_\mu+q_\mu)-\frac{i}{2}c_\text{SW}^{(0)}\sum\limits_\nu\sigma_{\mu\nu}(p_\nu-q_\nu)\right)+\mathcal{O}(a^2)\right)
\end{align}
Sandwiching its expansion in powers of $a$ with on-shell spinors $u$ and $\bar{u}$:
\begin{align}
\bar{u}(q)\Lambda^{a(0)}_\mu(p,q)u(p)=&-g_0T^a \bar{u}(q)\left(i\gamma_\mu+\frac{a}{2}\left(1-\csw^{(0)}\right)(p_\mu+q_\mu)\right)u(p)\nonumber\\
&+\mathcal{O}(a^2).
\end{align}
gives the condition $\cswzero=1$ to eliminate $\mathcal{O}(a)$ contributions\footnote{Technically $\cswzero=r$ where $r$ is the Wilson parameter.} .

At one-loop level the general form of the vertex function is 
\begin{align}
g_0^3\Lambda^{a(1)}_\mu = -g_0^3 T^{a} \left(\gamma_\mu F_1 + a \slashed{q} \gamma_\mu F_2 + a \gamma_\mu\slashed{p} F_3 +a(p_\mu+q_\mu)G_1  +a(p_\mu-q_\mu)H_1\right)
\end{align}
At $\mathcal{O}(a)$ $F_2$ and $F_3$ do not contribute on-shell and $H_1$ vanishes due to symmetry arguments \cite{Aoki:2003sj}. Sandwiching with on-shell spinors again gives
\begin{align}
&g_0^3\bar{u}(q)\left(\frac{a}{2}\csw^{(1)}(p_\mu+q_\mu)T^a + \Lambda^{a(1)}_\mu(p,q) \right)u(p)\nonumber\\
=&g_0^3\bar{u}(q)\left(i\gamma_\mu F_1+ \frac{a}{2}(p_\mu+q_\mu)(\csw^{(1)}-2G_1) T^a\right)u(p)+\mathcal{O}(p^2,q^2)+\mathcal{O}(a^2),
\end{align}
which results in the condition 
\begin{align}
\cswone=2G_1.
\end{align}
In Ref. \cite{Aoki:2003sj} it is pointed out, that $G_1$ can be easily extracted from the vertex function by
\begin{align}
G_1=-\frac18 \text{Tr}\left(\left(\frac{\partial}{\partial p_\mu}+\frac{\partial}{\partial q_\mu}\right)\Lambda^{a(1)}_\mu - \left(\frac{\partial}{\partial p_\nu}-\frac{\partial}{\partial q_\nu}\right)\Lambda^{a(1)}_\mu\gamma_\nu\gamma_\mu  \right)^{\mu\neq\nu}_{p,q\rightarrow 0}.
\end{align}
The six diagrams that contribute at one loop level are shown in figure 1. 
\begin{figure}
\centering
\includegraphics[scale=0.6]{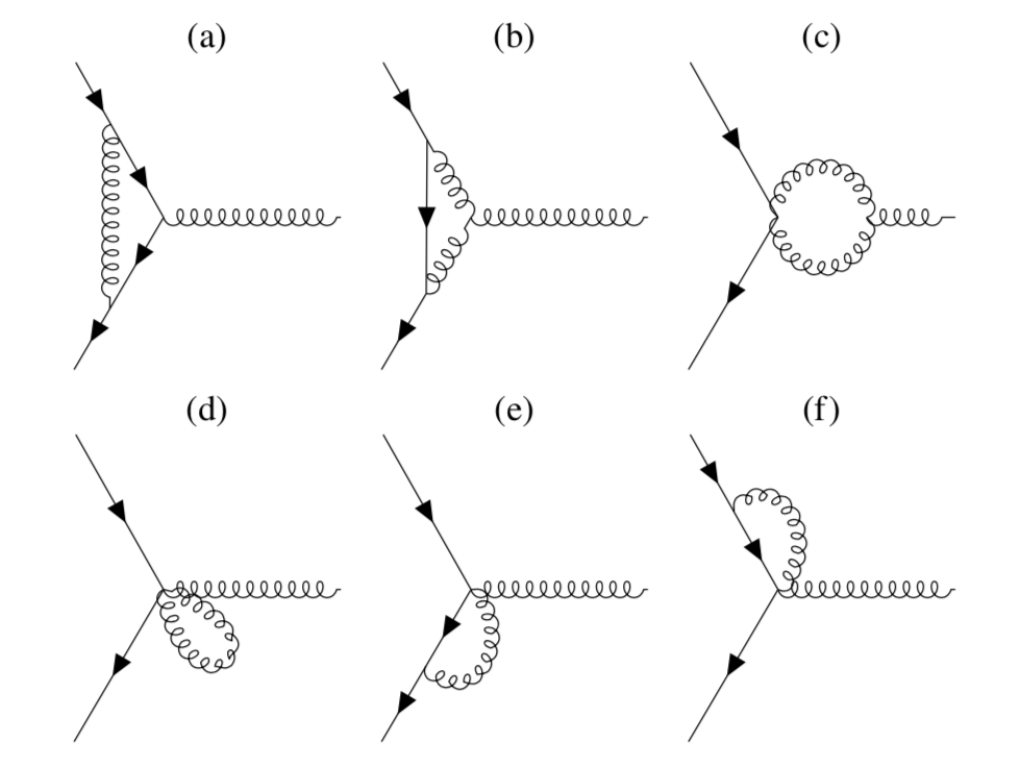}
\caption{The six one-loop diagrams contributing to the vertex function in lattice perturbation theory. }
\end{figure}

\begin{align}
{\Lambda^a_\mu}^{(1)}=\sum\limits_{i=a,...,f}{\Lambda^a_\mu}^{(1)(i)}=\sum\limits_{i=a,...,f}\intdk {I^a_\mu}^{(i)}(p,q,k).
\end{align}
All of them except the tadpole diagram (d) lead to IR-divergent integrals. To calculate them separately their divergence needs to be analytically split off and the remaining constant part calculated numerically. One possibility is to use a small fictitious gluon mass $\mu$  (as was done in \cite{Aoki:2003sj}). To this end an analytically solvable integrand with the same divergent behaviour $J$  is subtracted from the original integrand $I$:
\begin{align}
G_1^{(i)}&=\int\limits_{-\pi}^{\pi} \frac{d^4k}{(2\pi)^4}\left[I_\mu^{(i)}(k)-\theta(\pi^2-k^2)J_\mu^{(i)}(k,0) \right] + \frac{\Omega_3}{(2\pi)^4 }\int\limits_{0}^{\pi} J_\mu^{(i)}(k,\mu^2) k^3 dk
\end{align}
The Heaviside-function makes $J$ analytically solvable containing the non-zero gluon mass. 
As an example, let us consider diagram (c):
\begin{align}
{I^a_\mu}^{(c)}&=\sum\limits_{b,c}\sum\limits_{\nu,\rho} 
V_{2\,\nu\rho}^{bc}(p,q,q-p-k)\ G(k)\ G(p-q+k)V_{g3\,\nu\rho\mu}^{bca}(p-q+k,-k,q-p),\phantom{\sum\limits_\rho}\label{eq:1}
\end{align}
where $V_{2\,\mu\nu}^{ab}$, $V_{g3\,\mu\nu\rho}^{abc}$ and $G(k)$ are the lattice versions of the qqgg-vertex, the ggg-vertex and the gluon propagator respectively.
A possible analytically solvable integral can be obtained by expanding these to $\mathcal{O}(a)$ which gives: 
\begin{align}
\mathrm{J^{(c)}}&=\frac{3}{4}N_c\cswzero\frac{2\pi^2}{(2\pi)^4}\int\limits_0^\pi\frac{k^3}{(k^2+\mu^2)^2 }dk\\
&=\frac{3}{4}N_c\cswzero\frac{1}{16\pi^2}\left(\ln\left(\tfrac{\pi^2}{\mu^2}\right)-1\right)+\mathcal{O}(\mu^2)
\end{align}
Denoting the logarithmic divergence by $L=\frac{1}{16\pi^2}\ln\left(\tfrac{\pi^2}{\mu^2}\right)$ and adding the finite result from the $I-J$ integral results in the following contribution from diagram (c):
\begin{align}
2G_1^{(c)}=\frac92 L -0.0813095
\end{align}
The following table shows the contributions from all diagrams and how their divergent parts add up to zero. The far right column shows the values from \cite{Aoki:2003sj} for comparison. 
\begin{figure}[h!]
\centering
\begin{tabular}{|c|c|l|l|}
\hline
Diagram & Divergent part & Constant part & Aoki, Kuramashi \\
\hline
(a)&$-L/3 $ & 0.00457196 & 0.004572(2) \\
(b)&$-9 L/2$ & 0.0830768   & 0.08311(3) \\
(c)&$ 9 L/2$ & $-0.0813095$ & $-0.08133(3)$\\
(d)& 0  & 0.297394537 & 0.29739454(1)\\
(e)&$ L/6$ & $-0.0175746$ & $-0.017574(1)$ \\
(f)&$ L/6$ &$ -0.0175746$ & $-0.017574(1)$ \\
\hline 
Sum&0&$0.268588292$ & $0.26858825(1)$\\
\hline
\end{tabular}
\end{figure}
It is worth pointing out that calculating the finite sum of all diagrams directly instead of diagram by diagram can lead to a numerically more accurate result, as the discontinuity introduced through the Heaviside function in (\ref{eq:1}) causes  a slower convergence.

\section{Perturbative stout smearing and Wilson flow}
Stout smearing with smearing parameter $\rho$ of the link variable $U_\mu(x)$ is defined through
\begin{align}
U^{(1)}_\mu(x)=e^{i\rho Q_\mu(x)}U_\mu(x)
\end{align}
with
\begin{align}
Q_\mu(x)&=\frac 1{2i}\left( W_\mu(x)-\frac 13 \text{Tr}[W_\mu(x)]\right)\\
W_\mu(x)&=\sum\limits_{\nu\neq\mu} (P_{\mu\nu}^\dagger(x)-P_{\mu\nu}(x)),
\end{align}
where $P_{\mu\nu}(x)$ is the plaquette in the $\mu,\nu$-plane at lattice position $x$.
As $Q_\mu(x)$ is anti-hermitian, $U^{(1)}_\mu(x)$ is automatically in $SU(3)$ again and the smearing can be iterated leading to $U^{(n)}_\mu(x)$ after $n$ smearing steps. 

Wilson flow is the gradient flow of the Wilson action $\mathcal{S}_W$ and defined through the following differential equation: 
\begin{align}
\partial_tU_\mu(x,t)&=-g_0^2\left\{\partial_{x,\mu}\mathcal{S}_W[U(x,t)]\right\}U_\mu(x,t)=iQ_\mu(x,t)U_\mu(x,t)\\
U_\mu(x,0)&=U_\mu(x).
\end{align}
Therefore it is easily seen, that the Wilson flow is generated by infinitesimal stout-smearings. This means calculations done in the stout-smearing formalism with parameters $\rho$ and $n$ can be transformed into the Wilson flow formalism with flow time $t$ by performing the limit  $n\to \infty$, $\rho\to 0$ with $n\rho=t=\,$const. ($t$ in lattice units, i.e. $t/a^2$).

Our goal is to perform perturbative calculations of $\csw$  as sketched in section \ref{sect:2} including stout smearing and eventually Wilson flow. Therefore we need a perturbative expansion of the smearing.
\subsection{Leading order}
The un-smeared original link variable $U_\mu(x)$ has an expansion in terms of the gluon field $A_\mu(x)$:
\begin{align}
U_\mu(x)=1+ig_0T^aA^a_\mu(x)-\frac{g_0^2}{2}\frac 12\{T^a,T^b\}A^a_\mu(x)A^b_\mu(x)+\mathcal{O}(g_0^3)
\end{align} 
At leading order the smeared link variable $U^{(1)}_\mu(x)$ has a similar expansion with a modified gluon field $A^{a(1)}_\mu(x)$:
\begin{align}
U^{(1)}_\mu(x)&=1+ig_0T^a A^{a(1)}_\mu(x)+\mathcal{O}(g_0^2)\\
A_\mu^{a(1)}(x)&=A^a_\mu(x)+\rho(Q^a_\mu(x))^{\mathrm{LO}}
\end{align}
with
\begin{align}
(Q^a_\mu(x))^{\mathrm{LO}}=&\sum\limits_{\nu=1}^4\left(A_\nu^a(x)+A_\mu^a(x+\hat{\nu})-A_\nu^a(x+\hat{\mu})\right.\\
&\left.-A_\nu^a(x-\hat{\nu})+A_\mu^a(x-\hat{\nu})+A_\nu^a(x+\hat{\mu}-\hat{\nu})-2A_\mu^a(x)\right).
\end{align}	
After a Fourier transform $A^{a(1)}_\mu(k)$ can be expressed as \cite{Bernard:1999kc}: 
\begin{align}
A_\mu^{a(1)}(k)&=\sum\limits_\nu \left(f(k)\delta_{\mu\nu}-(f(k)-1)\frac{\hat{k}_\mu\hat{k}_\nu}{\hat{k}^2}\right)A^a_\nu(k)=:\sum\limits_\nu \tilde{g}_{\mu\nu}(\rho,k)A^a_\nu(k)
\end{align}
with $\hat{k}_\mu=2\sin(\tfrac 12 k_\mu)$ and $f(k)=1-\rho \hat{k}^2$. After $n$ smearing steps the overall structure remains the same except for the powers of $n$ 
\begin{align}
A_\mu^{a(n)}(k)&=\sum\limits_\nu\tilde{g}^n_{\mu\nu}(\rho,k)A^a_\nu(k)=\sum\limits_\nu \left(f(k)^n\delta_{\mu\nu}-(f(k)^n-1)\frac{\hat{k}_\mu\hat{k}_\nu}{\hat{k}^2}\right)A^a_\nu(k),
\end{align}
which in turn makes it easy to perform the aforementioned limit to the Wilson flow formalism:
\begin{align}
A_\mu^{a}(k,t)&=\sum\limits_\nu \left(e^{-t\hat{k}^2}\delta_{\mu\nu}-(e^{-t\hat{k}^2}-1)\frac{\hat{k}_\mu\hat{k}_\nu}{\hat{k}^2}\right)A^a_\nu(k,0)=:\sum\limits_\nu B_{\mu\nu}(k,t)A^a_\nu(k,0).
\end{align}

\subsection{Next-to-leading order}
At next-to-leading order in addition to the quadratic term involving the already known modified gluon field $A^{a(1)}_\mu(x)$ there is also an anti-symmetric part  $A_\mu^{ab(1)}(x)$:
\begin{align}
U^{(1)}_\mu(x)&=1+ig_0T^a A^{a(1)}_\mu(x)\nonumber\\
&-\frac{g_0^2}{2}\left(\frac 12 \{T^a,T^b\}A_\mu^{a(1)}(x)A_\mu^{b(1)}(x)+\frac 12  [T^a,T^b] A^{ab(1)}_\mu(x)\right)\nonumber\\
&+\mathcal{O}(g_0^3)
\end{align}
with
\begin{align}
A^{ab(1)}_\mu(x)&=4\rho\cdot \left[\frac 12 (Q^a_\mu(x))^{\mathrm{LO}}A_\mu^b(x)+(Q^{ab}_\mu(x))^{\mathrm{NLO}}\right].
\end{align}
Fourier transforming gives:
\begin{align}
A_\mu^{ab(1)}(k_1,k_2)=4i\rho \sum\limits_{\nu,\rho} g_{\mu\nu\rho}(k_1,k_2) A^a_\nu(k_1)A^b_\rho(k_2),
\end{align}
where
\begin{align}
g_{\mu\nu\rho}(k_1,k_2)
=&\delta_{\mu\nu}\sin(\tfrac 12( 2 k_{1\rho}+k_{2\rho}))\cos(\tfrac 12 k_{2\mu})\nonumber\\
-&\delta_{\mu\rho}\sin(\tfrac 12( 2 k_{2\nu}+k_{1\nu}))\cos(\tfrac 12 k_{1\mu})\nonumber\\
-&\delta_{\nu\rho}\sin(\tfrac 12( k_{1\mu}-k_{2\mu}))\cos(\tfrac 12( k_{1\nu}+k_{2\nu})).
\end{align}
After $n$ iterations we get
\begin{align}
A_\mu^{(n)[ab]}(k_1,k_2)=&
\sum\limits_\nu
\tilde{g}_{\mu\nu}(\rho,k_1+k_2)
A_\nu^{(n-1)[ab]}(k_1,k_2)
\nonumber\\&
+4 i \rho \sum\limits_{\nu,\rho} g_{\mu\nu\rho}(k_1,k_2) A^{(n-1)a}_\nu(k_1)A^{(n-1)b}_\rho(k_2)
\end{align}
 which we can express as a sum over the $A_\mu^{(m)}(x)$ from all previous smearing steps:
\begin{align}
A_\mu^{(n)[ab]}(k_1,k_2)=&
\sum\limits_{\nu\rho\sigma}
4 i \rho\,g_{\nu\rho\sigma}(k_1,k_2)
\sum_{m=0}^{n-1}
\tilde{g}^{n-m-1}_{\mu\nu}(\rho,k_1+k_2)
 A^{(m)a}_\rho(k_1)A^{(m)b}_\sigma(k_2)
\end{align}
And finally in the Wilson flow limit the sum turns into an integral over the flow time
\begin{align}
A^{[ab]}_\mu(k_1,k_2,t)=
\sum\limits_{\alpha\beta\nu\rho}B_{\mu\alpha}(k_1+k_2,t)\int\limits_0^t B_{\alpha\beta}(k_1+k_2,-t')4 i g_{\beta\nu\rho}(k_1,k_2)A^{a}_\nu(k_1,t')A^{b}_\rho(k_2,t')dt'
\,.
\end{align}
\section{Outlook}
In order to calculate diagram (d), which involves the qqggg-vertex, the smearing relation at next-to-next-to-leading order is also needed. There the expressions become rather lengthy. The next step will then be to insert the smearing relations into the expanded action to obtain the Feynman rules, which can only be partly compared to \cite{Horsley:2008ap}, as the ones coming from the clover term are not included there. \\

Another question we would like to investigate concerns the structure of the smearing expansion. Because the smeared link variable $U^{(1)}_\mu(x)$ is again in $SU(3)$ it is expected to have an expansion
\begin{align}
U^{(1)}_\mu(x)=e^{ig_1T^a\tilde{A}^a_\mu(x)}
\end{align}
with a modified (renormalised) coupling $g_1$.
However this expansion does not coincide with the expansion in $g_0$ and finding a link between the two may  help to simplify calculations.

\end{document}